\documentclass[a4paper,twoside,11pt]{article}
\usepackage{epsfig}
\def \pom {{\hspace{ -0.1em}I\hspace{-0.2em}P}}
\newcommand{\lsim}{\raisebox{-0.5mm}
{$\stackrel{<}{\scriptstyle{\sim}}$}}
\newcommand{\ffig}[4]{\begin{figure}[htbp]\vfill\begin{center}
\mbox{\epsfig{figure=#1,height=#2}}\caption{#3}\label{#4}
\end{center}\vfill\end{figure}}

\begin{document}
\noindent
DESY 96-019 \hfill ISSN 0418-9833\\
February 1996\\
\begin{center}
 \begin{Large}
 \begin{bf}
Soft Colour Interactions and the Diffractive
Structure Function\footnote{Presented by G.~Ingelman at Workshop on 
`HERA Physics -- Proton, Photon and Pomeron Structure', Durham, 
September 1995, to appear in the proceedings.}\\
 \end{bf}
 \end{Large}
 \vspace*{5mm}
 \begin{large}
A.~Edin$^a$, G.~Ingelman$^{a,b}$, J.~Rathsman$^a$\\
 \end{large}
\vspace*{3mm}
 {\it $^{a}$Dept. of Radiation Sciences, Uppsala University, 
 Box 535, S-751 21 Uppsala, Sweden}\\
 {\it $^{b}$Deutsches Elektronen-Synchrotron DESY, Notkestrasse 85,
 D-22603 Hamburg, Germany}\\
\vspace*{5mm}
\end{center}
\begin{quotation}
\noindent
{\bf Abstract:}
We discuss our model for soft colour interactions and present results
on the diffractive structure function $F_2^D(\beta,x_\pom,Q^2)$
and inclusive transverse energy flows, which agree with available HERA 
data.
\end{quotation}

The observation of rapidity gap events at HERA \cite{ZEUS,H1} has led
to a renewed interest in diffractive phenomena and how they can be
understood within QCD. In \cite{EIR} we presented a new model for 
diffractive hard scattering based on a mechanism for soft colour
interactions (SCI). 
The starting point is the normal DIS parton level
interactions, with perturbative QCD corrections based on standard
first order matrix elements and conventional leading log parton
showers for higher order parton emission. This gives a state of
colour-ordered partons that will have further non-perturbative
interactions to produce the final hadron state. The novel feature in
our model is the introduction \cite{EIR} of random soft colour
interactions corresponding to non-perturbative gluon exchange between
these partons. This changes the colour structure and thereby affects
the hadronic final state when a conventional hadronisation model such
as the Lund string model \cite{Lund} is applied. 

Rapidity gaps may then arise when a gap at the parton level is not 
spanned by a string. In particular, when the hard process starts with a
gluon from the proton, leaving a colour octet remnant and a colour
octet hard scattering system, a soft colour exchange between the two
octet systems can give two colour singlets separated in rapidity 
\cite{EIR}. Large forward rapidity gaps are here favoured by the large
momentum of the remnant, in particular at small-$x$. Our model
\cite{EIRplan} is implemented in the Monte Carlo (MC) {\sc Lepto} 6.4
producing complete events.

To compare our model with experimental data on rapidity gap events we
consider the diffractive structure function $F_2^D(\beta,x_\pom,Q^2)$
defined by \cite{IP}
\begin{equation}
\frac{d\sigma^{D}}{d\beta dx_\pom dQ^2}=
\frac{4\pi\alpha^2}{\beta Q^4}\left[1-y+\frac{y^2}{2}\right]
F_2^D(\beta,x_\pom,Q^2)
\end{equation}
(i.e. assuming single photon exchange and neglecting $F_L$). This
inclusive quantity contains the dependence on the main variables 
$\beta \simeq Q^2/(Q^2+M_X^2)$, $x_\pom \simeq (Q^2+M_X^2)/(Q^2+W^2)$ 
and the usual DIS momentum transfer squared $Q^2$. The acceptance
corrected data from H1 \cite{H1} are compared in Fig.~1 to our model
results obtained by selecting MC events with rapidity gaps similar to
the H1 definition (i.e.\ no energy in $\eta_{max}<\eta<6.6$ where
$\eta_{max}<3.2$).

The model is generally in good agreement with the data. It has a
tendency to be below the data at large $Q^2$, possibly due to
slightly too much parton radiation. The
$\beta$-dependence seems to be the same in the model and the data and
thereby the $M_X$ dependence is also basically correct. The
$x_\pom$-dependence in the model may be slightly steeper than in the
data. Fitting a universal $x_\pom^{-a}$-dependence we get $a=1.5\pm
0.2$ from our model to be compared with the H1 result $a=1.2\pm 0.1$
\cite{H1} and ZEUS results $a=1.3\pm 0.1$ \cite{ZEUS} and 
$a=1.5\pm 0.1$ \cite{ZEUS2} using different methods.
However the result obtained in the model depends significantly on which 
$x_\pom$-bins
that are included in the fitting procedure. This originates from the
model not quite giving a straight line in this plot, but having a 
tendency for a curvature with smaller slope at small $x_\pom$ and 
larger slope at large $x_\pom$. With improved experimental data this 
aspect of the model can be tested.

\ffig{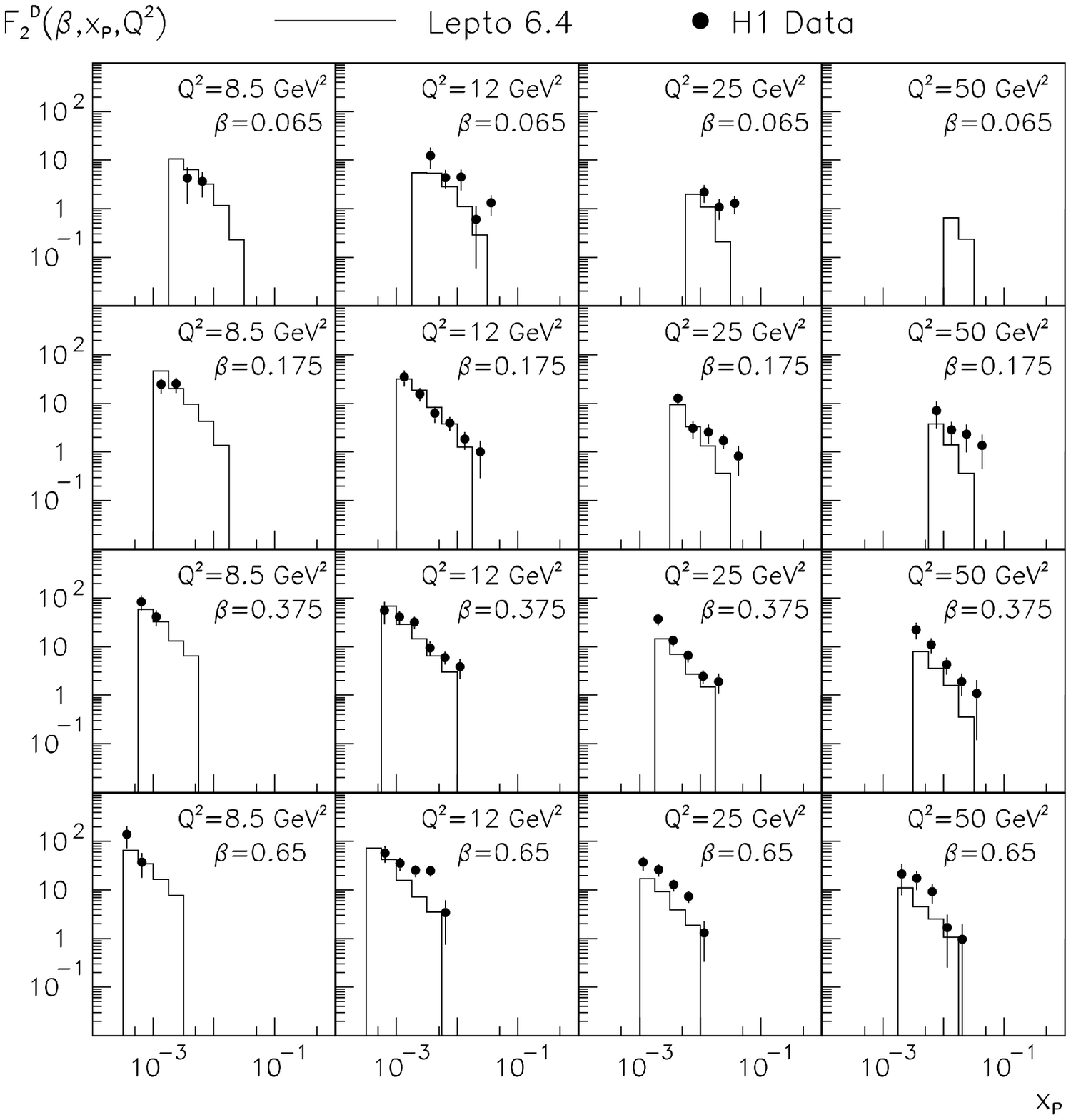}{160mm}{
The diffractive structure function $F_2^D(\beta,x_\pom,Q^2)$ 
from our soft colour interaction model (histograms) compared to the H1 data points 
\protect\cite{H1}.}{fig2}

Our model is meant to be applicable for DIS in general and should
therefore also be compared with normal DIS events. The observed large
forward transverse energy flow in an inclusive event sample requires a
substantial energy and particle production and is thereby `orthogonal'
to forward rapidity gaps. In Fig.~\ref{fig3} we compare our model
result with the H1 data \cite{H1eflow}. The agreement is quite good
except for the smallest $x$-values where the model is below the data in
the central region of the hadronic cms.

The possibility to obtain rapidity gaps depends on to which extent
there are gaps at the parton level, i.e. on the amount of parton
emission in the forward region. In our model we use the first order QCD
matrix elements for the primary emission and therefore the treatment of
their soft and collinear divergences is of importance. We have replaced
the conventional requirement $m_{ij}^2>y_{cut}W^2$ on any pair $ij$ of
partons with an advantageous `mixed' scheme using cuts in
$\hat{s}=(p_1+p_2)^2$ and $z=p_1P/qP$ which handles initial state
collinear singularities better \cite{EIRplan}. Numerically we use
$\hat{s}_{\min}=1$ GeV$^2$ and $z_{\min}=0.01$. 

\ffig{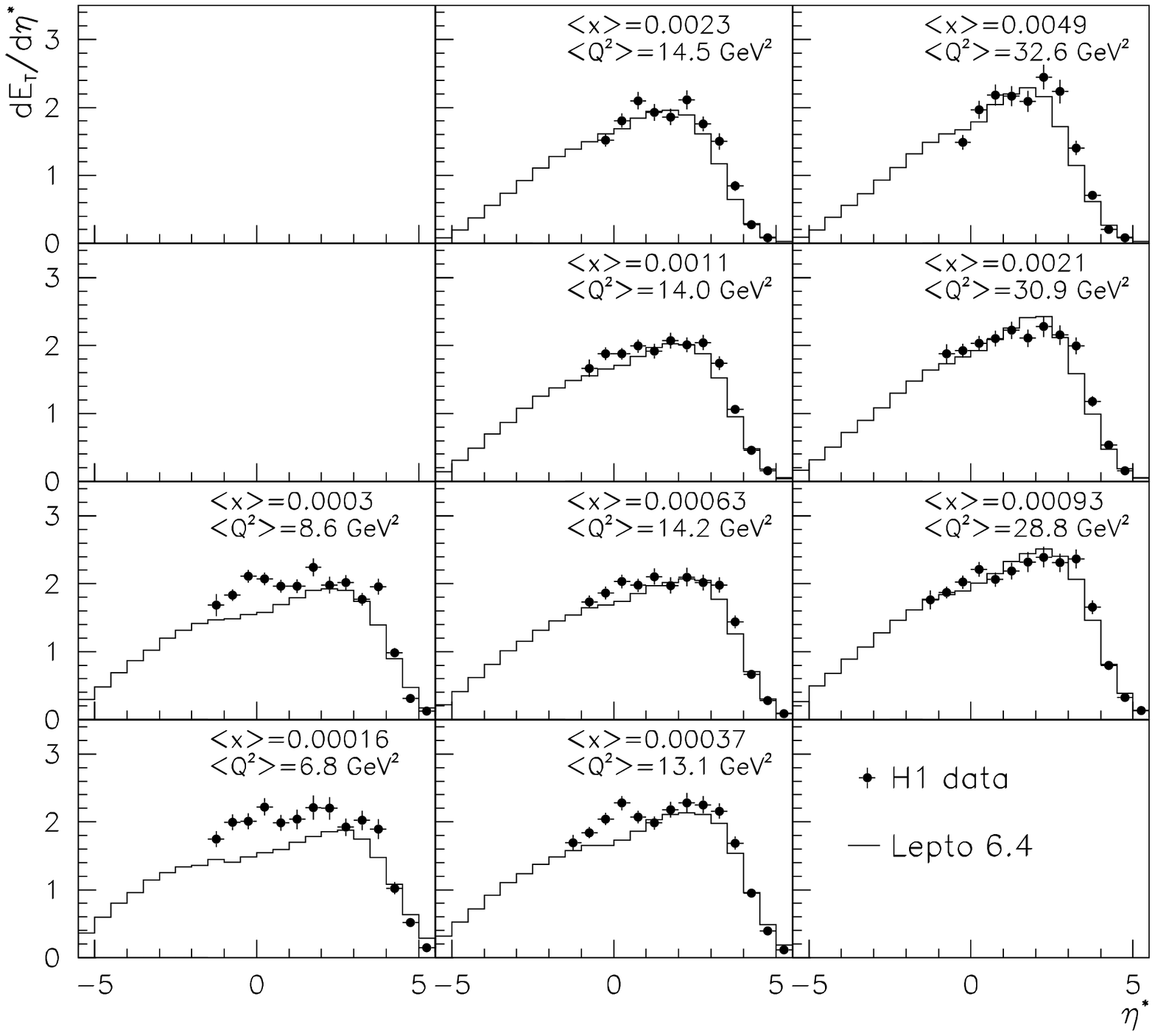}{140mm}{
Transverse energy flow versus pseudorapidity $\eta^*$ in the hadronic 
cms from our model (histograms) compared to H1 data points 
\protect\cite{H1eflow}.}{fig3}

Higher order parton emissions are taken into account approximately
through parton showers developing from the incoming and scattered
parton. The initial state shower is most important for the forward 
rapidity region. 
In our model we are using the conventional GLAP evolution
scheme \cite{GLAP} summing leading $\log{Q^2}$ terms. Terms with
$\log(1/x)$ are neglected in GLAP, but will become important at small
$x$ and should then be resummed as in the BFKL equation \cite{BFKL}.
Therefore GLAP evolution should no longer be valid at some small $x$.
Recent studies \cite{LDCM} based on the CCFM equation \cite{CCFM} 
which sums both leading $\log{Q^2}$ and $\log(1/x)$ terms
suggests that this happens only at very small $x$, below the region 
$10^{-4}\lsim \,x\, \lsim 10^{-2}$  where the rapidity gap events are
observed. Thus, there is no strong reason that the GLAP formalism
should not be applicable for our purposes. 

One may worry that unsuppressed parton emission will destroy the gaps
\cite{LL}. The cut-off in parton virtuality, defining the borderline to
the non-perturbative region, is a regulator of this. We have previously
\cite{EIR} used the value 2 GeV, but in our improved model
\cite{EIRplan} returned to 1 GeV, which has been standard and conforms
with e.g. LEP analyses. Thus, our model works with a conventional
unsuppressed GLAP parton shower. 

Of importance for the gap rate is also the fluctuations in the initial 
parton emission \cite{EIR}. Although, one may expect that a GLAP parton
shower gives a fair mean description of events, there is no guarantee
that it accounts properly for fluctuations. Larger fluctuations of the
number of emitted gluons would increase the rate of gap events  and also 
increase the inclusive forward energy flow due to `downwards' and 
`upwards' fluctuations, respectively.

In summary, our model based on non-perturbative soft colour
interactions can explain important features 
of both rapidity gap events and normal DIS interactions at HERA.

\end{document}